\documentclass[5p]{elsarticle}

\usepackage{amsmath}				
\usepackage{hyperref}				
\usepackage{graphicx}				
\usepackage{subfig}					
\newcommand{\mat}[1]{\mathbf{#1}}		
\renewcommand{\vec}[1]{\mathbf{#1}}	
\usepackage{appendix}
\usepackage{float}

\usepackage[table]{xcolor}				
\definecolor{lightgray}{gray}{.8}			
\definecolor{lightblue}{rgb}{.827,.875,.933}	
\usepackage{lscape}		
\usepackage{textcomp}		

\newlength{\SMALLitemsep}
\newlength{\SMALLparskip}
\begin{document}

\title{Network structure of inter-industry flows}

\author[BU,SFI]{James McNerney\corref{cor}}
\ead{jamesm@bu.edu}
\author[TowsonU,IIASA]{Brian D. Fath}
\ead{bfath@towson.edu}
\author[UMaastricht,IIASA]{Gerald Silverberg}
\ead{gerald.silverberg@merit.unimaas.nl}
\address[BU]{Department of Physics, Boston University, Boston, MA 02215, USA}
\address[SFI]{Santa Fe Institute, 1399 Hyde Park Road, Santa Fe, NM 87508, USA}
\cortext[cor]{Corresponding author}
\address[UMaastricht]{UNU-MERIT, Maastricht University, Keizer Karelplein 19, 6211 TC  Maastricht, The Netherlands}
\address[TowsonU]{Department of Biological Sciences, Towson University, 8000 York Road, Towson, MD 21252, USA}
\address[IIASA]{International Institute for Applied Systems Analysis, Schlossplatz 1, A-2361, Laxenberg, Austria}


\begin{abstract}
We study the structure of inter-industry relationships using networks of money flows between industries in 20 national economies. We find these networks vary around a typical structure characterized by a Weibull link weight distribution,  exponential industry size distribution, and a common community structure. The community structure is hierarchical, with the top level of the hierarchy comprising five industry communities: food industries, chemical industries, manufacturing industries, service industries, and extraction industries.
\end{abstract}

\begin{keyword}
industrial network \sep input/output table \sep money flows \sep national accounting \sep macroeconomics
\end{keyword}


\maketitle


\section{Introduction}
Goods in an economy are produced by a network of industries, where each industry produces goods by combining the output of others. The structure of this network may provide clues to how economies function and eventually shed light on how economies change over time. While direct data on physical production flows between industries are unavailable, data on money flows are. This study presents some initial findings about the structure of this money flow network, with a particular emphasis on patterns that are shared across economies and can serve as targets for statistical physics models.

Money flows fall into a number of large categories of transactions, such as output, consumption, income, and investment. Also included are the somewhat smaller (though still large) flows between industries. National accounting provides a system for cataloguing these money flows. Although national accounting does not use network terminology to describe these flows, they are naturally expressed in these terms, with links representing flows and nodes representing industries or sectors. Here, we focus on a subset of money flows, those within the business sector, which comprises the industries of an economy (Fig. \ref{fig:economy_diagram}). The resulting web of industrial trading is therefore not a closed network but an open one, with flows entering and exiting from outside.

Our data comes from input/output (I/O) tables, which are part of the national accounting data compiled by national statistical agencies. The I/O tables are quite similar to adjacency matrices, with several additional rows and columns added to account for boundary flows, changes to stocks, and special categories of goods, as well as separate tables to account for import flows.

A few studies have already applied network approaches to the I/O tables. These studies are roughly divided between empirical studies of structure \cite{Slater1977,Slater1978,Aroche-Reyes2003,Carvalho2007} and theoretical models of dynamics \cite{Leontief1986,Carvalho2007,Bloechl2011}. The structure studies suggest the existence of clustering among industries. Carvalho \cite{Carvalho2007} further finds an asymmetry between in-flows and out-flows of industries that implies an asymmetry between industries as providers of goods and users of goods. While different industries tend to require similar numbers of input goods, they may provide inputs to either many or few other industries -- showing that some industries are general purpose providers while others are specialists. The models of dynamics have focused on the role that network structure may play in economy-wide fluctuations, by modeling how shocks propagate through the web of industries. However, despite previous work, many basic properties of these networks remain uninvestigated. 

This paper is a step towards eventually building network models of economies. We begin by explaining some basic principles of national accounting and the measurement basis for money flows, since these concepts may not be common knowledge among physicists. We then analyze industrial networks in terms of topology, flow size distribution, industry size distribution, and community structure. Our findings suggest that industrial networks have rich structure that is susceptible to analysis using complex systems approaches.

The paper is organized as follows. In Section \ref{sec:national_accounting} we explain the principles of national accounting. In Section \ref{sec:description_of_data} we describe the data set. In Section \ref{sec:network_characteristics} we discuss the topology, flow size distribution, industry size distribution, and community structure of the industrial networks in our data set. In Section \ref{sec:discussion} we discuss our results.

\section{National accounting}\label{sec:national_accounting}
\begin{figure*}[t!]
\textsf{\textbf{a}}\\ \includegraphics[width=.7\textwidth]{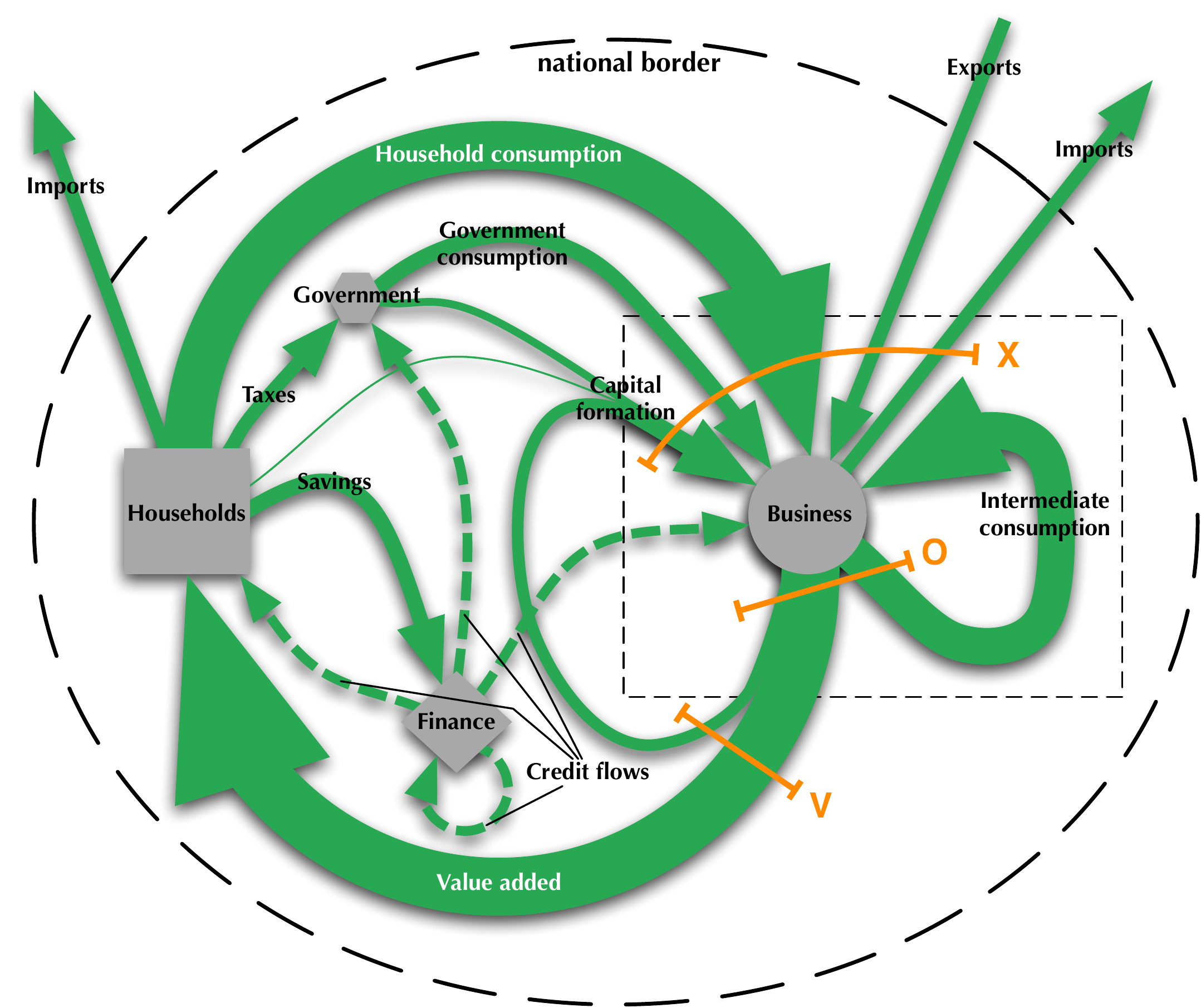}
\textsf{\textbf{b}} \includegraphics[width=.3\textwidth]{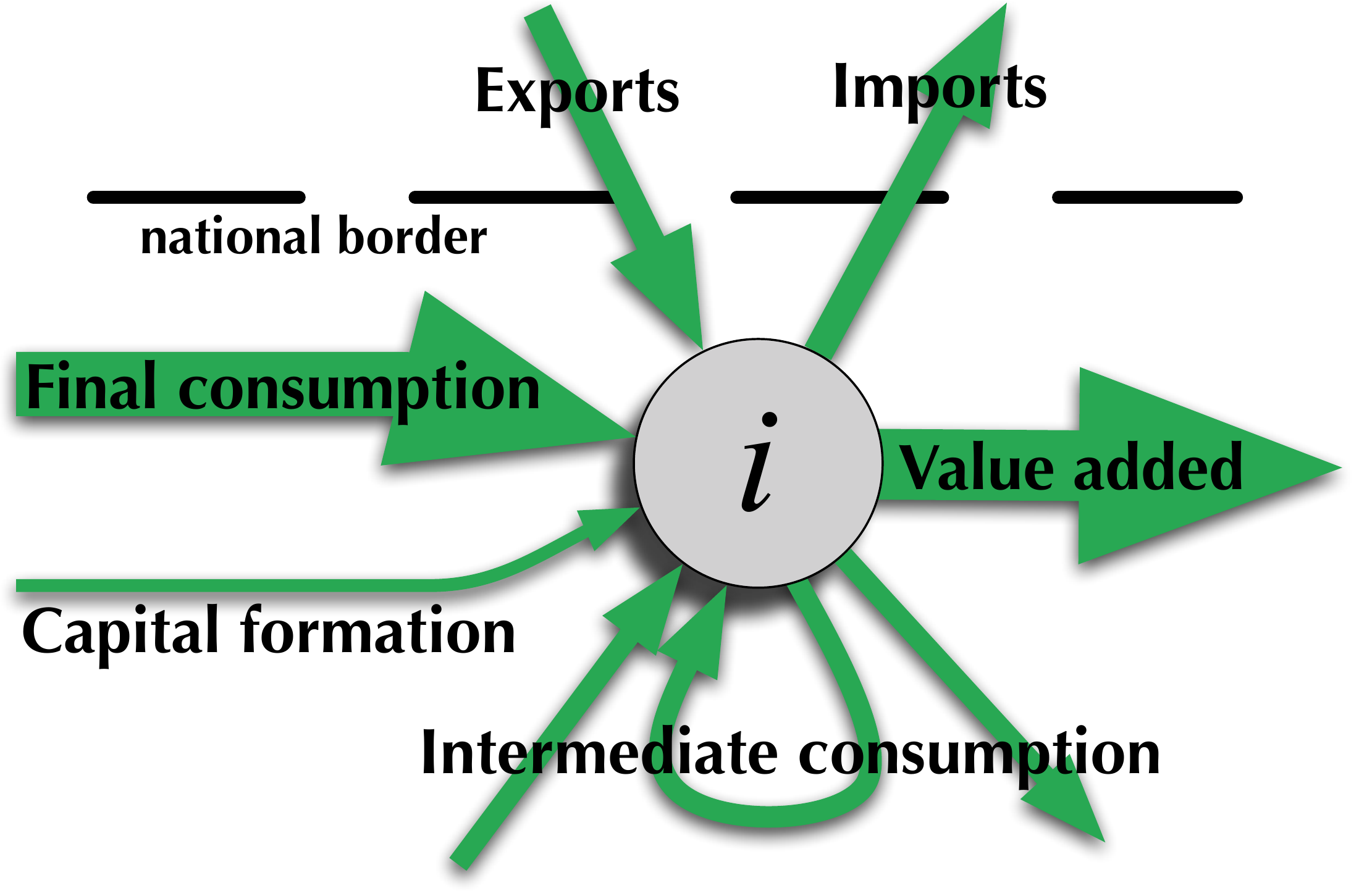}
\caption[Sectoral level money flows]{(a) Flows of money between sectors in an economy. The dashed box indicates the scope of the I/O tables. The 3 ``gates'' X, O, and V show approximately where the 3 methods for measuring GDP capture flows to compute GDP. (See Eq. \eqref{GDP_equation}.) Gate X corresponds to the ``expenditure approach'', gate O to the ``output approach'', and gate V to the ``value-added approach''. Although the finance sector is shown apart from the business sector, the I/O tables do include inferred service fee flows between finance and industries of the business sector. Credit flows and deposits, however, are outside the scope of the I/O tables. For clarity, several features are not shown: credit flows from non-finance sectors, business taxes, government subsides, transfer payments, government imports, government self-flows, investment to foreign countries. Not all interest flows are shown, but can be inferred from credit flows. One of the five SNA sectors, non-profits, is also not shown for clarity. (b) Flows through a particular industry of the business sector.}
\label{fig:economy_diagram}
\end{figure*}
Measurement of money flows involves substantially more complications than measurement of other kinds of network flows, such as energy, information, or air passenger traffic, due to the many categories of transactions that are separately accounted for and the conventions of national accounting. In this section we briefly describe national accounting methods for quantifying these flows.

First it is useful to describe the general structure of economies; this broader context helps make sense of the logic behind industry network data. Economies are composed of five ``institutional sectors'' or simply ``sectors'' for short: households, non-financial business, financial business, government, and non-profits. (Fig. \ref{fig:economy_diagram}a) The largest money flows are ``household consumption'' -- purchases of business sector goods and services by the household sector -- and ``value added''. Value added partly corresponds to purchases of household sector labor by the business sector, though it contains other components as well, as we discuss further on. Household consumption and value added collectively are referred to as the ``circular flow'' by economists and constitute the backbone of sectoral money flow structure. Note that the circular flow in this sense refers only to the monetary aspect of the economy. Biophysical flows, which also have a circular component, are maintained by boundary flows from free energy to wastes that have no monetary analog.

Next in size is ``intermediate consumption'' by the business sector. Intermediate consumption represents purchases made by industries for goods produced by other industries. Whereas household consumption goods are intrinsically desirable -- a bottle of cola, say -- goods purchased for intermediate consumption are not, but are inputs required to produce intrinsically desirable goods -- e.g., carbonated water, syrup, and glass. Intermediate consumption is recorded in input/output tables, which are a part of the accounting system used by national statistical agencies to record economic activity.

Capital purchases are an important exclusion from transactions classified as intermediate consumption. Capital purchases are purchases for goods that aid the production of other goods \emph{and} can be used repeatedly over time -- a bottling machine, say. Goods are classified as capital goods when they can be used repeatedly for more than one accounting period, usually one year. Most input/output tables only record the industry selling a capital good and not the industry buying it. Thus, instead of a full adjacency matrix of capital purchases, input/output data usually only records a vector of capital revenues received by each industry.

The transactions underlying money flows in the network are compiled on an ``accrual basis''. Under this accounting system, revenues are recognized when they are earned by the transfer of goods or the performance of a service. Expenses are recognized when the associated revenues are earned. To see how this affects the recording of money flows, consider a car maker purchasing steel, producing a car, and selling it over some period of time. Under an accrual basis, the accounts of the car maker -- and those of the automotive industry in the input/output tables -- will record sales revenue being received when the car is transferred to the consumer, even if the full purchase price is not paid immediately. At the same time, the steel expense will be matched to the car that it helped produce, even though such expenses were actually incurred earlier. The alternative method of recording transactions is ``cash-flow basis'' accounting, in which transactions are recognized when money is paid or received. Accrual basis accounting can be thought of as a pseudo-goods tracking approach, because it follows the movement of goods rather than the movement of money.\footnote{An important special case of the distinction between accrual basis and cash-flow basis transactions is depreciation flows. Depreciation in accounting is the assignment of portions of a fixed expense to multiple time periods. Depreciation transactions are recorded as though the depreciable asset is consumed over time. The consumption of a depreciable asset is thus recognized as a transaction many times throughout the depreciable lifetime of the asset, even though no literal cash flow occurs.}

Money flows within the full sector-level network are not conserved for at least two reasons. First, money may disappear from accidental loss or destruction. Second and more importantly, money is regularly created and destroyed by the financial sector. National accounting does enforce a virtual conservation law, though, through the use of balancing items, which are accounting entries that are calculated as the difference between other accounting entries. In the I/O tables, the balancing item is value-added, which is calculated as the difference between total sales by the business sector and intermediate consumption sales. Value-added ``measures the value created by production'' \cite{SNA2008} and encompass all forms of personal income -- employee compensation, interest, dividends, and rent, as well as certain kinds of taxes and depreciation.

Finally, though it is not essential to our purpose of analyzing industry networks, it is useful to understand how GDP is calculated and how it relates to industry networks. Exploiting the conservation enforced by the definition of value-added, one can equate money flows in and out of the business sector:
\begin{align}\label{business_throughflow}
{\small
\begin{array}{r}
\text{value added}\\
+ \text{intermediate consumption}\\
+ \text{imports}\\
+ \text{business taxes}
\end{array}
=
\begin{array}{l}
\text{intermediate consumption}\\
+ \text{household consumption}\\
+ \text{government consumption}\\
+ \text{capital formation}\\
+ \text{exports}\\
+ \text{subsidies}
\end{array}}
\end{align}
Or, by rearranging terms,
\begin{align}\label{GDP_equation}
{\small
\begin{array}{r}
\text{value added}\\
+ \text{business taxes}\\
- \text{subsidies}
\end{array}
=
\begin{array}{l}
\text{household consumption}\\
+ \text{government consumption}\\
+ \text{capital formation}\\
+ \text{exports}\\
- \text{imports}
\end{array}
\equiv \text{GDP}}.
\end{align}
The left hand side represents the ``income approach'' to calculating GDP, in which forms of income are summed. The right hand side represents the ``expenditure approach'' to calculating GDP. By using the identity ``$\text{value added} = \text{gross output} - \text{intermediate consumption}$'', a third approach can be derived -- the ``output approach'' -- where value added is calculated as the difference between all business sales and intermediate goods sales. All three approaches are used by statistical agencies to validate GDP calculations. They also provide equivalent intuitive interpretations of GDP as a measure of total income, a measure of total expenditures, and as the net output of the business sector.
\begin{table*}[t!]
\center
\caption{Country data statistics.}
\rowcolors{2}{}{lightblue}
\small
\begin{tabular}{lcccc}
\hline\hline
\rowcolor{lightgray}
\textbf{Country}
&\textbf{Year}
&\textbf{Num. industries in data}
&\textbf{Fraction self-flows}
&\textbf{Completeness}
\rule[-2mm]{0pt}{4ex}\\
\hline
Australia	&1994-95	&38	&0.215	&0.999\\
Brazil	&1996	&30	&0.240	&0.998\\
Canada	&1997	&34	&0.232	&0.969\\
China	&1997	&38	&0.238	&0.943\\
Czech Republic	&1995	&40	&0.292	&0.965\\
Denmark	&1997	&39	&0.179	&0.957\\
Finland	&1995	&35	&0.274	&0.977\\
France	&1995	&39	&0.285	&0.776\\
Germany	&1995	&36	&0.228	&0.995\\
Greece	&1994	&36	&0.168	&0.929\\
Hungary	&1998	&36	&0.237	&1.000\\
Italy	&1992	&37	&0.247	&0.854\\
Japan	&1995	&40	&0.219	&0.818\\
Korea	&1995	&39	&0.253	&0.888\\
Netherlands	&1995	&38	&0.260	&0.907\\
Norway	&1997	&40	&0.204	&0.999\\
Poland	&1995	&35	&0.270	&0.998\\
Spain	&1995	&39	&0.225	&0.961\\
United Kingdom	&1998	&40	&0.286	&0.949\\
United States	&1997	&39	&0.238	&0.994\\
\hline\hline
\end{tabular}
\label{tab:country_data_statistics}
\end{table*}

\section{Description of data}\label{sec:description_of_data}
Our data comes from 1997 I/O tables produced by the Organization for Economic Cooperation and Development (OECD) \cite{OECDweb}. The tables describe intermediate consumption flows in 20 countries (not all OECD members) between 41 industries. The I/O data were initially collected by national statistical agencies, who followed country-specific practices for partitioning the business sector into industries and measuring flows. The OECD converted country data sets into a uniform 41-industry system to make international comparisons possible.

The countries are listed in Table \ref{tab:country_data_statistics} and the industries in Table \ref{tab:industry_statistics}. One industry, ``Private households with employed persons'', was excluded from analysis because it was poorly represented in the data, with only 3 out of 20 countries (Australia, Japan, and Korea) providing any data for it. This industry represents the production activity of cooks, butlers, chauffeurs, gardeners, nannies, etc. and does not make a significant contribution to flows in any of the 3 countries where data is available.

Because the I/O tables of individual countries differed in both the number of industries and the boundaries between them, the translation step between the national system and SNA involved undesired splits and mergers that affect the size of flows and industries. When an industry defined by the source country overlapped two or more of the industries defined by the OECD, the OECD was forced to choose which OECD industry to assign the source industry to. As a result, some industries in the OECD data represent more than their intended scope of production activities, while others represent less. In many instances, such mergers caused entire industries to be completely subsumed under other industries. Table \ref{tab:country_data_statistics} lists the number of industries represented after all mergers are taken into account.

\section{Network characteristics}\label{sec:network_characteristics}
\subsection{Notation}
Let $\mat{A}$ be the adjacency matrix for the money flows between industries. An element $A_{ij}$ denotes the flow from industry $j$ to industry $i$:
\begin{align}
A_{ij} = \text{flow from $j$ to $i$.}
\end{align}
Self links, representing payments of an industry to itself, are permitted.

\begin{figure}[t!]
\center
\includegraphics[width=.3\textwidth]{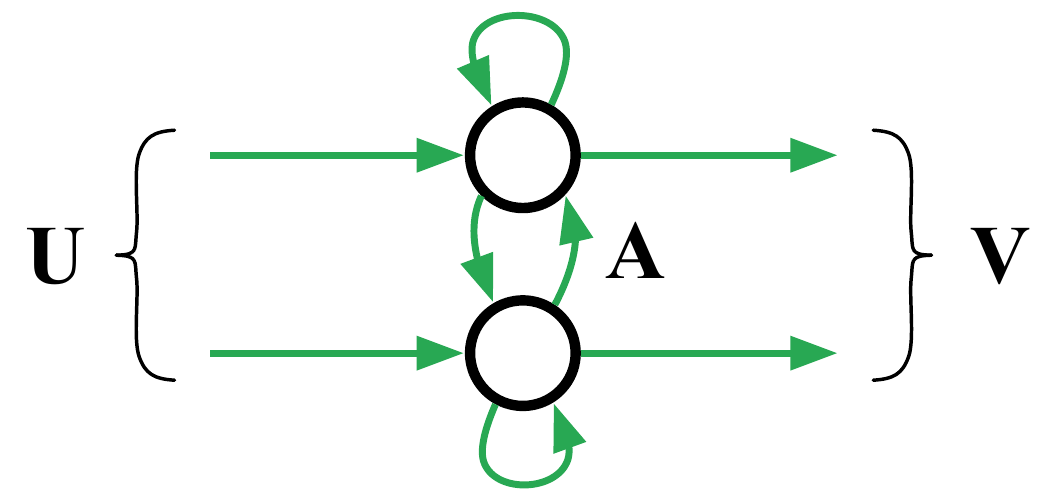}
\caption{Simplified networks structure and notation.}
\label{fig:toy_industrial_network}
\end{figure}

In addition to flows between nodes, an industrial network has in-flows entering the network from outside, and out-flows exiting the network (Fig. \ref{fig:toy_industrial_network}.) As explained in Section \ref{sec:national_accounting}, the in-flows correspond to final consumption, capital purchases, and export revenues. The out-flows correspond to value added and import expenditures. Let the sum over in-flows to each industry be denoted by the \emph{in-vector} $\vec{U}$, and let the sum over out-flows to each industry be denoted by the \emph{out-vector} $\vec{V}$:
\begin{align}
U_i &= \text{flow from outside to $i$}\\
V_i &= \text{flow from $i$ to outside}
\end{align}

Flow is conserved at all nodes because of the definition of value added, as described in Section \ref{sec:national_accounting}. At each node $i$, flow in equals flow out. Borrowing from ecology, we will refer to the flow into/out of node $i$ as its \emph{throughflow}, $T_i$ \cite{Fath2001}:
\begin{align}\label{throughflow}
T_i \equiv U_i + \sum_j A_{ij} &= \sum_j A_{ji} + V_i.
\end{align}
Summing over all nodes, the throughflow of the whole business sector is
\begin{align}
\Omega \equiv \sum_i T_i &= \sum_i U_i + \sum_{ij} A_{ij} = \sum_{ij} A_{ij} + \sum_i V_i.
\end{align}
This equation is the same as Eq. \eqref{business_throughflow}, with $\sum_{ij} A_{ij}$ corresponding to intermediate consumption and the other terms corresponding to $\sum_i V_i$ or $\sum_i U_i$. In economic terms, the total throughflow $\Omega$ represents gross output (the total of all sales by the business sector) plus imports and business taxes.

\subsection{Topology}
At the level of aggregation used in our data, industrial networks are nearly complete graphs, typically with more than 90\% of all possible flows having non-zero weight. (Table \ref{tab:country_data_statistics})

The high degree of completeness is only a feature of highly aggregated I/O tables. Carvalho, studying use tables\footnote{Use tables are a related data set that shows the expenditure of each industry on individual commodities. Use tables are similar to I/O tables and are used in their construction.} with approximately 500 industries, notes that the network is only 18\% complete at that level of aggregation \cite{Carvalho2007}.

\subsection{Flow weight distribution}
The magnitudes of money flows in different countries differ because they are expressed in different currencies and their economies vary in size. To make flow weights comparable across countries, we normalize them by the total throughflow of the country:
\begin{align}
a_{ij}^{c} \equiv \frac{A_{ij}^c}{\Omega^{c}},
\end{align}
where $\Omega^{c}$ is the throughflow of country $c$ and $a_{ij}^{c}$ is the normalized flow weight of country $c$.

\begin{figure}[t]
\includegraphics[width=.5\textwidth]{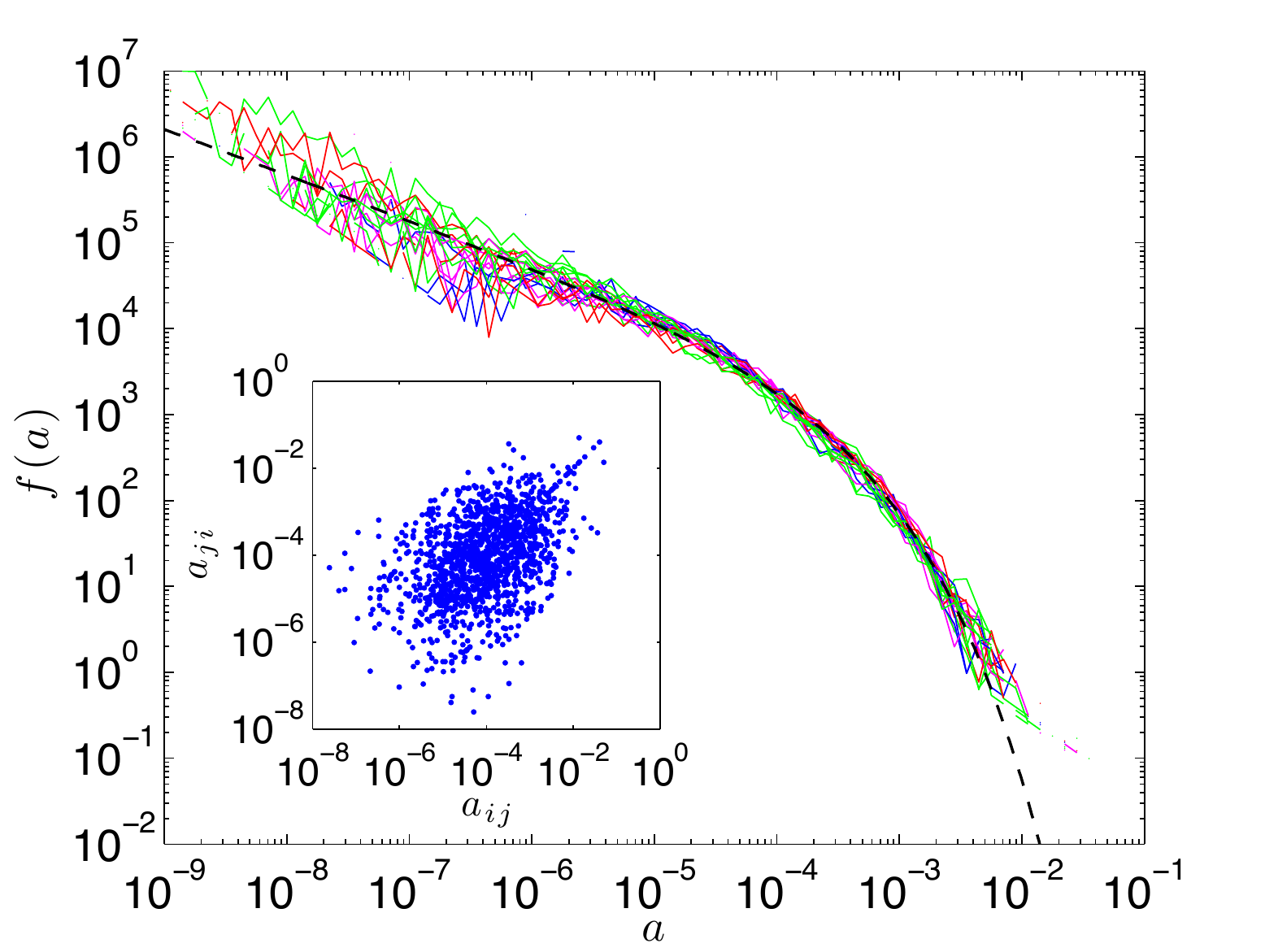}
\caption[Inter-industry flow weight distribution]{(Color online) Weight distributions of 20 countries studied. The dashed line is the best fit Weibull distribution to the pooled data from all 20 countries. Inset: $a_{ij}$ v. $a_{ji}$ for Spain.}
\label{fig:weight_distribution}
\end{figure}

The distributions of the normalized flow weights for all 20 countries are shown in Fig. \ref{fig:weight_distribution}. These distributions cover a wide range, with largest and smallest weights separated by 5 to 8 orders of magnitude, depending on the country. The flow weight distribution is heavy-tailed and shows significant curvature on log-log axes. It behaves very similarly for different countries throughout much of its range. At lower weights, the various country distributions diverge from each other to some extent.

The weight distributions are similar to both the Weibull,
\begin{align}
f(a) = \frac{k}{\lambda} \left( \frac{a}{\lambda} \right)^{k-1} \exp \left[ -\left(\frac{a}{\lambda}\right)^k \right]
\end{align}
and lognormal distributions,
\begin{align}
f(a) = \frac{1}{\sqrt{2\pi s^2}} \frac{1}{a} \exp\left[- \frac{(\ln a - m)^2}{2s^2} \right].
\end{align}
These two distributions are frequently difficult to distinguish in empirical data.\cite{Kundu2004}. A standard method for choosing the better fit between them is to compare the log-likelihoods from maximum likelihood fits of each distribution, accepting the distribution with the higher log-likelihood \cite{Kundu2004,Kundu2006,Kim2008}. Results are shown in Table \ref{tab:Weibull_v_lognormal}. Out of 20 countries, 11 are better described by a Weibull distribution and 9 by a lognormal. We also run a pooled regression under the assumption that the data follow approximately the same distribution. The pooled regression favors the Weibull and is shown as the dotted line in Fig. \ref{fig:weight_distribution}. In addition, two other factors favor the Weibull.
First, most countries do not show clear evidence of non-monotonic behavior, which would occur under a lognormal. Finland and Hungary are exceptions, showing a small amount non-monotonicity. Second, the Weibull tends to overestimate the occurrence of the smallest flows, while the lognormal tends to underestimate it. It is more likely that the smallest flows would be underrepresented in the data due to incomplete sampling rather than being overrepresented.

Because the network is simultaneously directed and nearly complete (at this level of aggregation), almost every flow $a_{ij}$ in the network has a reciprocating flow $a_{ji}$ of non-zero weight. The inset of Fig. \ref{fig:weight_distribution} plots weights against reciprocating weights for the United States IO network, with similar results for other countries. The correlation between off-diagonal elements is low (with typical correlation coefficients in the range  $\rho = 0.1$ to $0.4$). In many cases, a flow is several orders of magnitude larger or smaller than the reciprocating flow, indicating a high degree of asymmetry in the network. This is not surprising, since for most pairs of transacting industries, one industry is primarily the supplier and the other primarily the user.

\begin{figure}[t!]
\center
\includegraphics[width=.48\textwidth]{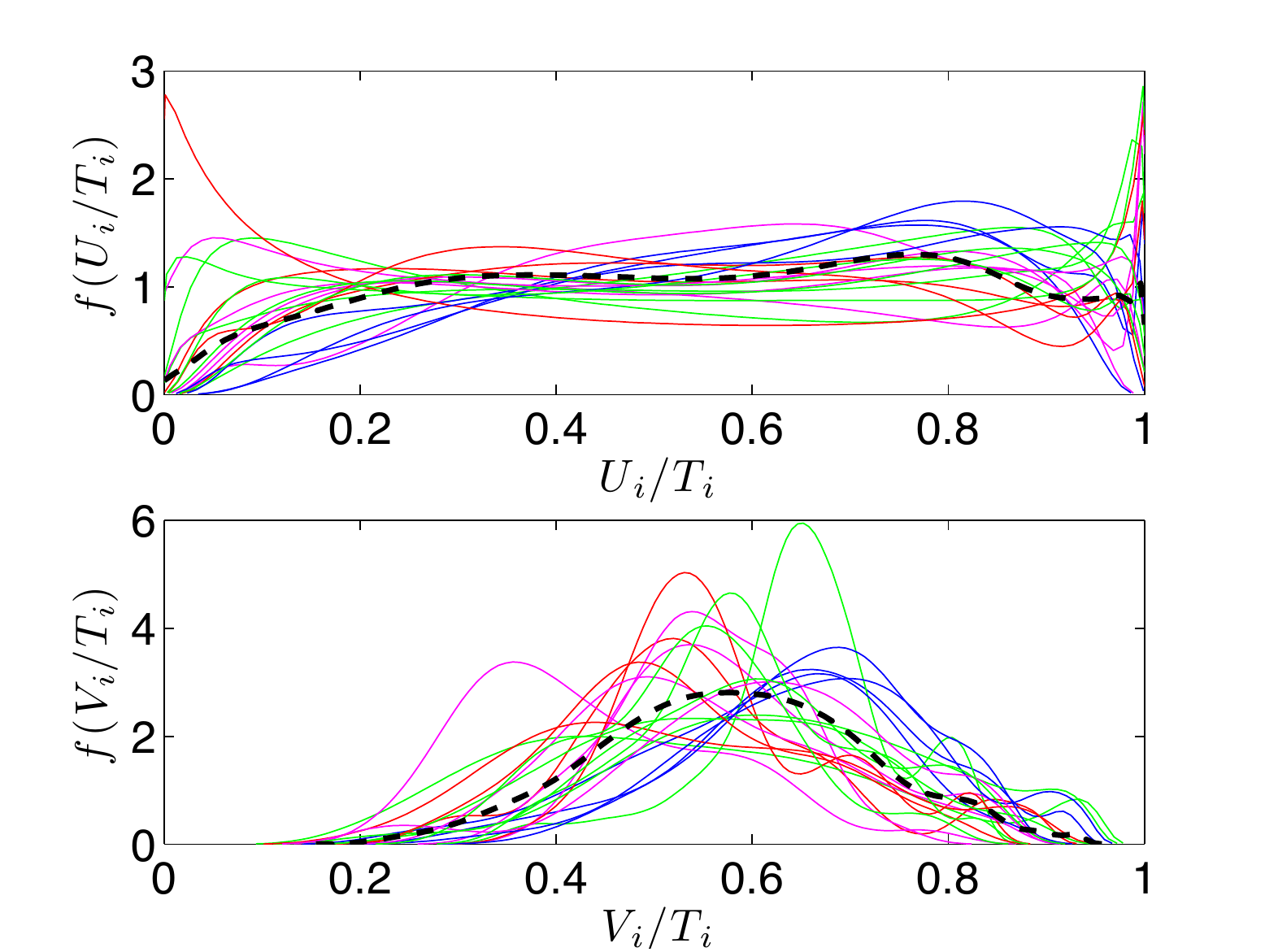}
\caption[Density of $U_i/T_i$ and $V_i/T_i$]{(Color online) Density of $U_i/T_i$ and $V_i/T_i$. Country lines (solid) were estimated used kernel density smoothing. The dashed line represents the pooled data.}
\label{fig:uv_distributions}
\end{figure}

The external flows, $U_i$ and $V_i$, between an industry $i$ and other sectors of the economy are generally much larger than flows between $i$ and other industries, and are comparable in size to the whole throughflow $T_i$. In Fig. \ref{fig:uv_distributions}, we plot the densities of $U_i/T_i$ and $V_i/T_i$. The first quantity is the fraction of money in-flows received from final consumption sales, sales of capital goods, and exports. (That is, all non-intermediate categories of receipts.) The second quantity is the fraction of money out-flows paid to value-added and imports (all non-intermediate categories of expenditures.) The density of $U_i/T_i$ is spread out across the whole interval $[0,1]$. This mainly reflects the large variation among industries in how directly they service final consumption, which is the most important component of $U_i$. In contrast, the density of $V_i/T_i$ is peaked, roughly around 0.6. This means that industries are more similar with respect to how much they spend on payments to the household sector than in how much they receive from it. This suggests that while industries differ significantly in where they lie on production chains, they have somwhat similar labor needs in monetary terms.

\subsection{Node throughflow distribution} \label{sec:throughflow_distribution}
Node strength generalizes the concept of node degree to weighted networks. Since the network is directed, each node $i$ has both an in-strength and an out-strength, defined as the sum of either the in-flows or out-flows incident on $i$. These sums are equal in this network due to flow conservation, so there is only quantity to keep track of, which we refer to as the \emph{throughflow} $T_i$ of node $i$. (Eq. \eqref{throughflow}.) As was done for link weights, we normalize node throughflows to render them comparable between countries:
\begin{align}\label{eq:normalized_thruflow}
t_i^c &\equiv \frac{T_i^c}{\Omega^c}.
\end{align}
The quantity $t_i$ measures the size of industry $i$ as the fraction of money flowing through industry $i$.
\begin{figure}[t]
\includegraphics[width=.5\textwidth]{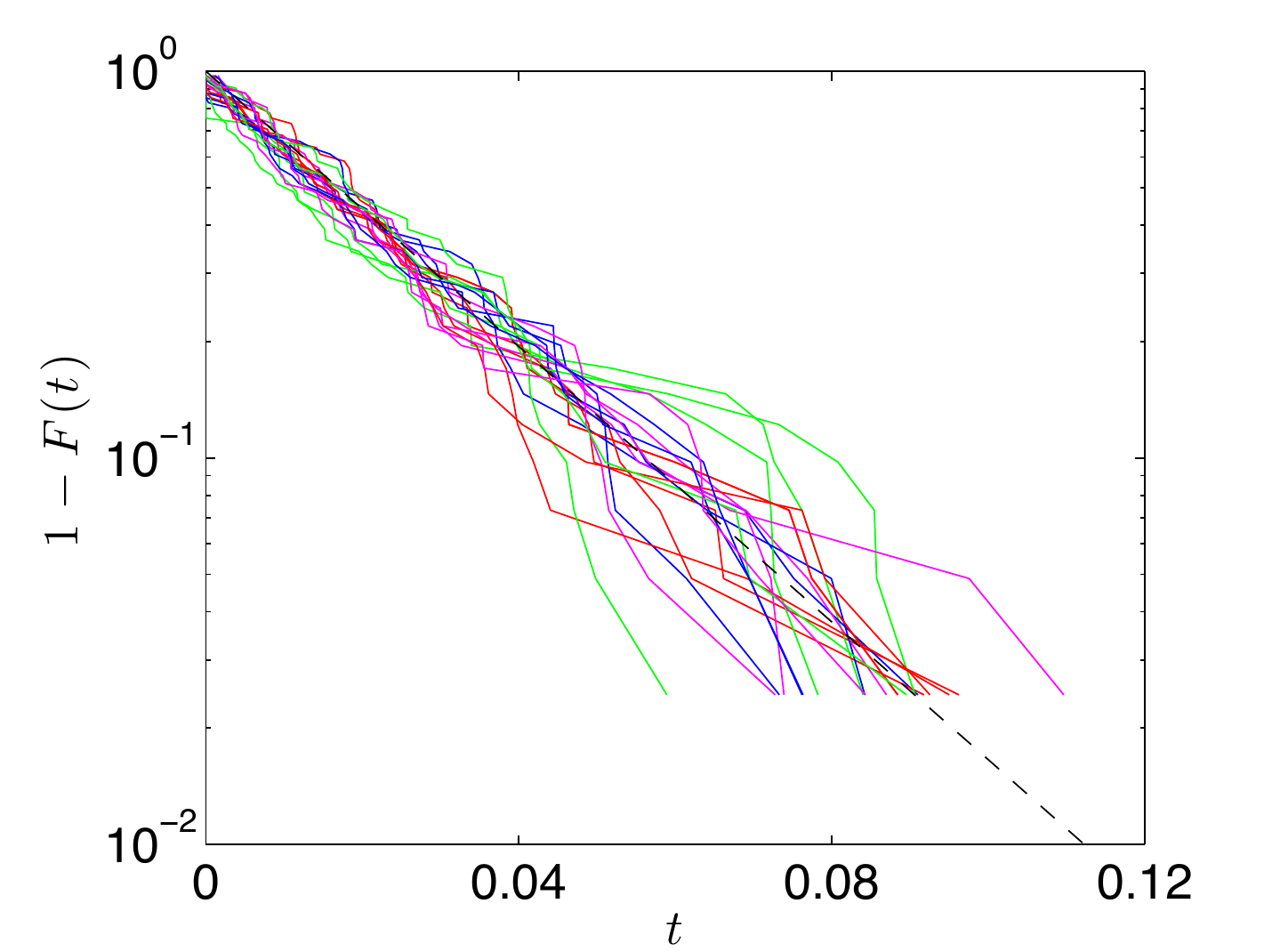}
\caption[Industry throughflow distribution]{(Color online) The throughflow distributions of all 20 countries studied.}
\label{fig:throughflow_distribution}
\end{figure}

The throughflow distributions of all 20 countries are shown in Fig. \ref{fig:throughflow_distribution}. The distribution is similar from country to country and is approximately exponential. 

Table \ref{tab:industry_statistics} shows the sizes of the 40 industries recognized in the OECD data. Under the OECD's partitioning of industries, the five largest industries are
\begin{itemize}
\setlength{\itemsep}{\SMALLitemsep}%
\setlength{\parskip}{\SMALLparskip}
\item wholesale and retail trade
\item construction
\item real estate activities
\item food, beverages, and tobacco
\item public administration and defense.
\end{itemize}
The industries most likely to export are
\begin{itemize}
\setlength{\itemsep}{\SMALLitemsep}%
\setlength{\parskip}{\SMALLparskip}
\item office, accounting, and computing machinery
\item aircraft and spacecraft
\item radio, television, and communication equipment
\item building and repairing of ships and boats
\item motor vehicles, trailers, and semi-trailers.
\end{itemize}
Unsurprisingly, the least likely to export are
\begin{itemize}
\setlength{\itemsep}{\SMALLitemsep}%
\setlength{\parskip}{\SMALLparskip}
\item real estate
\item health and social work
\item public administration and defense
\item education
\item construction,
\end{itemize}
all industries whose products are not easily traded across national borders. The industries receiving the most revenue from final demand are quite similar:
\begin{itemize}
\setlength{\itemsep}{\SMALLitemsep}%
\setlength{\parskip}{\SMALLparskip}
\item public administration and defense
\item education
\item health and social work
\item construction
\item real estate.
\end{itemize}
The industries least likely to receive revenue from final demand are
\begin{itemize}
\setlength{\itemsep}{\SMALLitemsep}%
\setlength{\parskip}{\SMALLparskip}
\item iron \& steel
\item non-ferrous metals
\item mining and quarrying
\item other non-metallic mineral products
\item rubber and plastic products.\footnote{See Chenery \& Watanabe \cite{Chenery1958} for a classification of industries based on the fraction of revenues from intermediate sales and the fraction of expenditures on intermediate goods. They use the first fraction to measure how ``final'' versus ``intermediate'' an industry is. They use the second to determine whether an industry is ``primary'' or ``manufacturing''. Using these two dimensions, they classify industries into four rough categories.}
\end{itemize}

\subsection{Community structure}\label{sec:community_structure}
In addition to knowing the statistics of flows and industry sizes, we would like to know whether industries cluster in any particular way. Such clusters are usually referred to as ``communities''. Many methods exist for finding communities in networks \cite{Porter2009,Fortunato2010}; here, we apply the method of modularity optimization \cite{Newman2004,Newman2010}. Modularity maximization involves searching for partitions of the network into communities that yield high values of the \emph{modularity} $Q$ over all possible partitions of the network. Since our network is directed, we use the directed generalization of modularity \cite{Leicht2008},
\begin{align}\label{modularity_function}
Q(c_1,\ldots,c_n) = \frac{1}{m} \sum_{ij} \left( a_{ij} - \frac{\hat{s}_i \check{s}_j}{m} \right) \delta(c_i, c_j),
\end{align}
Here, $c_i$ is the community that node $i$ belongs to, $m = \sum_{ij} a_{ij}$ is the total weight of all edges, and $\hat{s}_i = \sum_j a_{ji}$ and $\check{s}_i = \sum_j a_{ij}$. The Dirac delta function $\delta(k,l) = 1$ if $k=l$ and 0 otherwise. The modularity gives the total weight of edges within communities minus the expected weight under a null model. The modularity function scores a given partition of the nodes into groups; the task then is to search over the many possible partitions of the network and find the one with the highest score. In practice, the number of partitions is usually extremely large, so that only a small fraction can be examined directly. This has led to many proposals for algorithms that attempt to search the space of partitions efficiently for high values of $Q$ rather than find the global maximum \cite{Fortunato2010,Good2010}.


Recent work has shown that the modularity function $Q$ admits a large number of high-scoring partitions that are not necessarily similar \cite{Good2010}. As a result, different searches may arrive at different high-scoring partitions. Deterministic algorithms in particular are problematic because they fail to show the many alternative partitions. To address this problem, we use a stochastic search algorithm based on simulated annealing that returns a different high-scoring partition in each run. We repeat the algorithm many times, collect an alternative partition from each run, and compare them to test their robustness from run to run.

Specifically, we use the following simple procedure. For each country, we run the simulated annealing algorithm 100 times and extract 100 high-modularity partitions. From these partitions we produce a \emph{coclassification matrix} \cite{Sales-Pardo2007} with elements $p_{ij} \in [0,1]$ equal to the frequency with which node $i$ is grouped with node $j$. If certain nodes or groups of nodes are frequently grouped together, they will appear as blocks of high frequencies in the coclassification matrix; if the groups are highly variable, then no particular part of the matrix will accumulate a high value.

For the purpose of community finding, we set self-flows $a_{ii}$ of industries to zero, since these flows may reduce the resolution of the method. This happens because including self-flows increases $m$ in Eq. \eqref{modularity_function}, decreasing the null model ``penalty term'' $\hat{s}_i\check{s}_j/m$. This makes mergers between communities that we would like to distinguish more favorable, since it is then easier for a link between two industries to exceed the null model penalty term. A potential drawback of excluding self-flows is that if there are industries that should be classified as singleton communities, our method will not find them, because the associated term $a_{ii} - \hat{s}_i\check{s}_i/m$ in Eq. \eqref{modularity_function} can only contribute negatively to $Q$. However, in return we gain the benefit of more effectively resolving communities between two or more industries. This tradeoff is acceptable, since the communities we are interested in are \emph{inter}-industry ones. In fact, we find similar results whether self-flows are excluded or not, though we only show the results based on excluding self-flows.



\begin{figure*}[t!]
\begin{tabular}{cc}
\begin{minipage}{.25\textwidth}
\textsf{\textbf{a}} \textsf{Australia}\\
\includegraphics[width=1\textwidth]{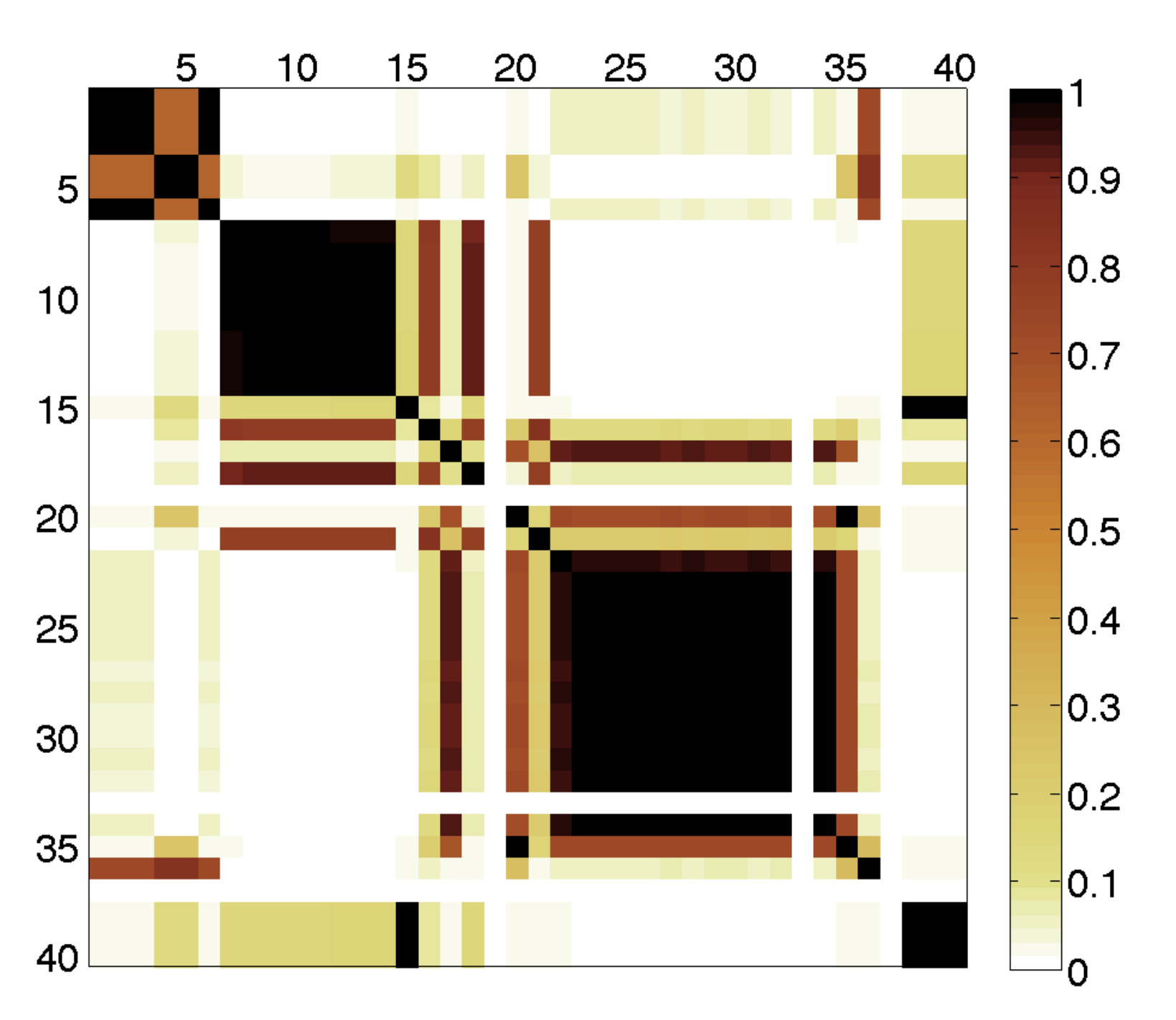}\\
\textsf{\textbf{b}} \textsf{China}\\
\includegraphics[width=1\textwidth]{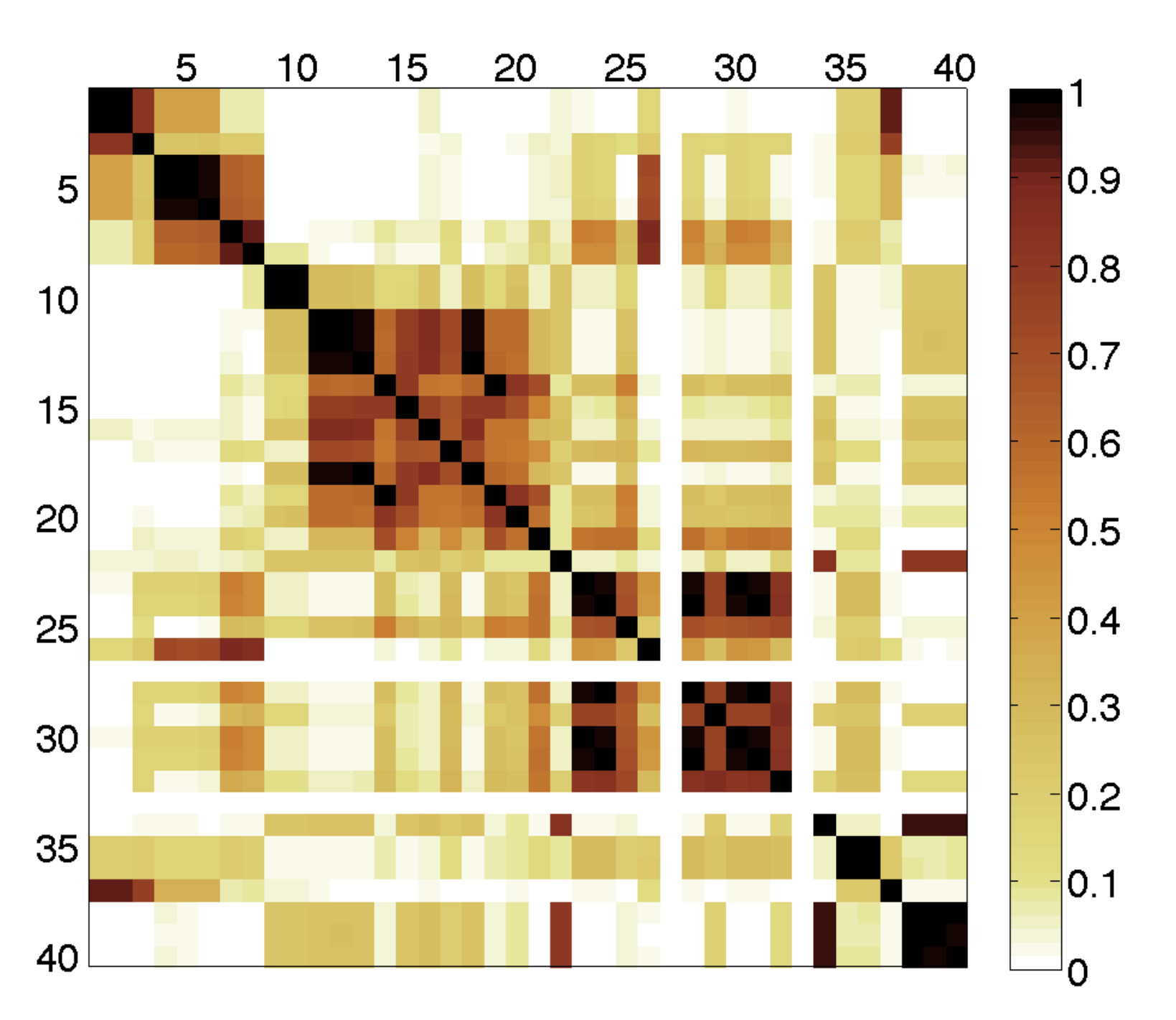}\\
\textsf{\textbf{c}} \textsf{United States}\\
\includegraphics[width=1\textwidth]{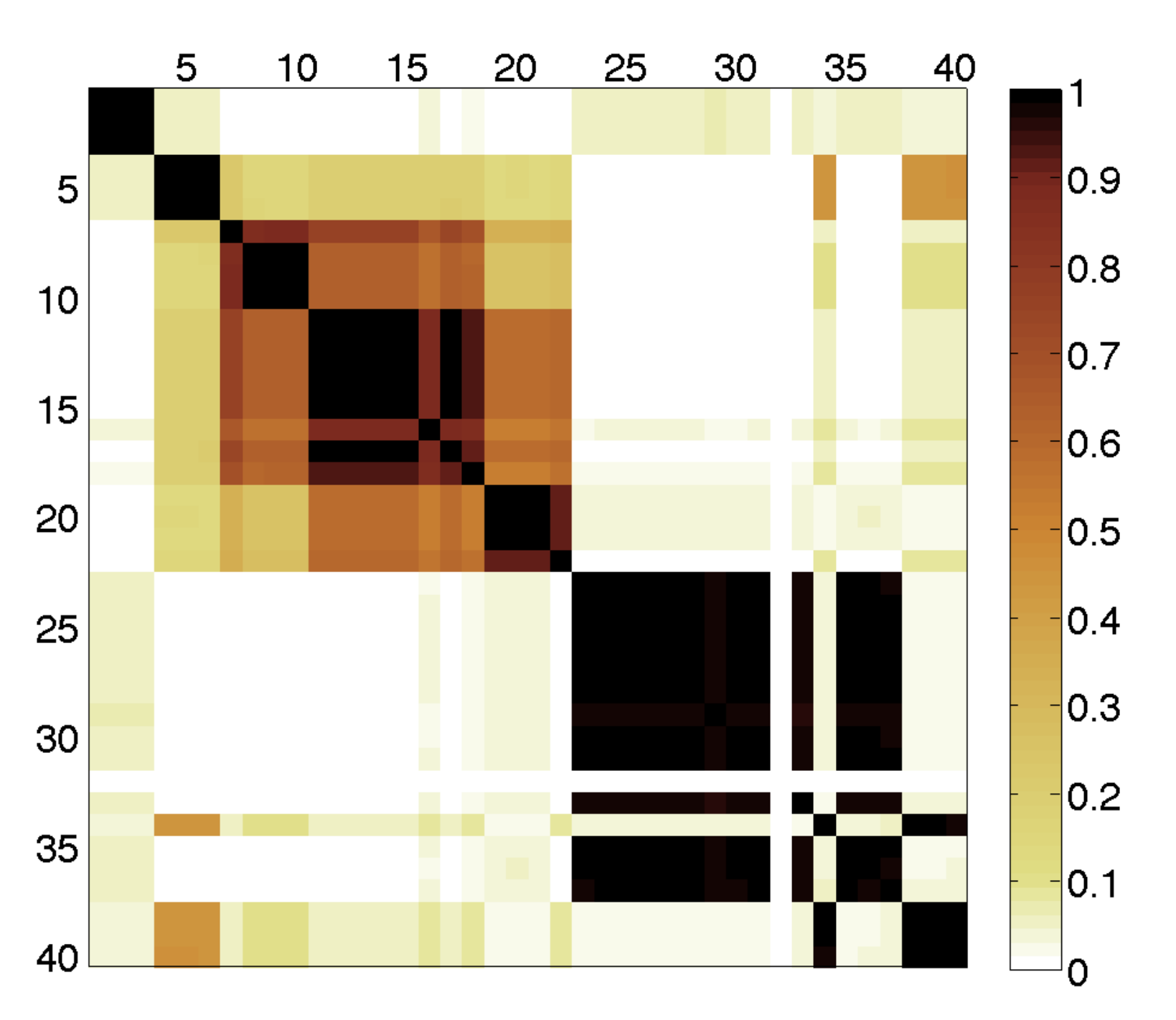}
\end{minipage}
&
\begin{minipage}{.75\textwidth}
\textsf{\textbf{d}} \hspace{105pt} \textsf{All 20 countries}\\
\includegraphics[width=1\textwidth]{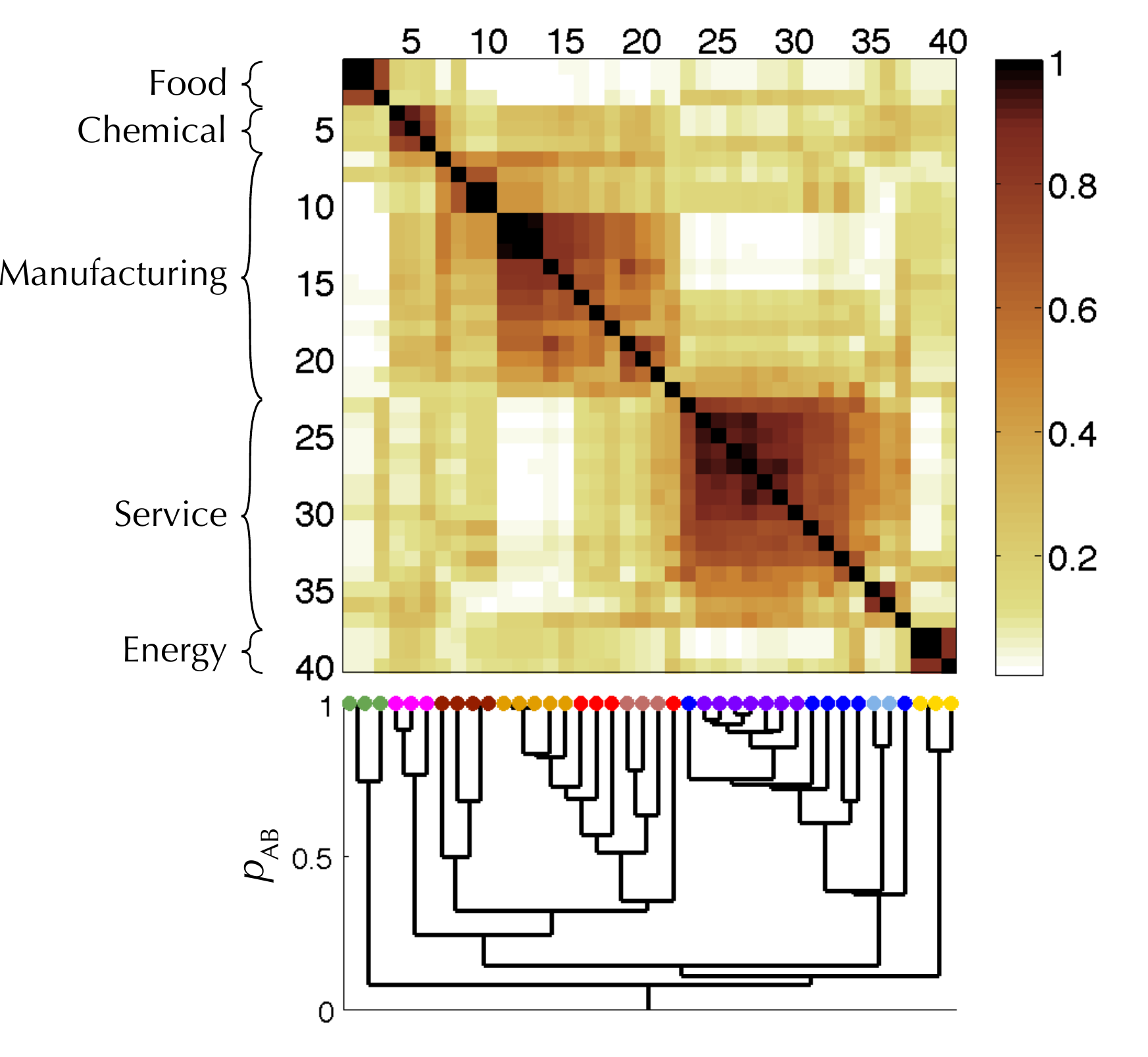}
\end{minipage}
\end{tabular}
\caption[Coclassification matrices for industry communities]{(Color online) Coclassification matrices (CCMs) giving the probability of two industries being grouped in the same community. Rows and columns correspond to the 40 economic industries in Table \ref{tab:industry_statistics}. \textbf{a}, \textbf{b}, and \textbf{c} CCMs for Australia, China, and United States. \textbf{d} Average CCM of all 20 countries in Table \ref{tab:country_data_statistics}, and dendrogram showing results of hierarchical clustering. The vertical axis of the dendrogram measures clustering probabilities $p_{AB}  = 1 - d_{AB}$.}
\label{fig:coclassification_matrix}
\end{figure*}

Figures \ref{fig:coclassification_matrix}a-c show the coclassification matrices for Australia, China, and the United States. These figures show the level of variation possible within countries from one simulated annealing run to another. Although both the communities and their stability varied somewhat from country to country, different countries nevertheless tended toward similar groupings corresponding to food industries (rows/columns 1-3), chemical industries (4-6), manufacturing industries (7-22), service industries (23-38), and energy industries (39-41). Unsurprisingly, industries had a higher tendency to transact with other industries of similar type.

To study this common tendency more closely, we constructed the average CCM of all 20 countries. The result is another CCM (Fig. \ref{fig:coclassification_matrix}d), whose $i$-$j$th element now indicates the frequency with which industries $i$ and $j$ were grouped together out of 2000 search runs (100 per country). Overall, the five-way grouping above performs well as a coarse-grained description of the community structure.

Going beyond this quick description, we can also study the matrix in Fig. \ref{fig:coclassification_matrix}d for clues of hierarchical community structure \cite{Good2010,Sales-Pardo2007}. Such structure arises in the CCM because industries with ambiguous community membership may switch back and forth across a community boundary between different runs of the search algorithm.

For example, the ``transport and storage'' industry may be grouped with service industries in one run, and with energy industries in another. The two runs may be different runs for the same country or for two different countries, as in the case of Fig. \ref{fig:coclassification_matrix}d. An industry that switches back and forth between one group and another will appear ``smeared'' across both groups. This indeed occurs for ``transport and storage'' ($i=34$). Other industries that show this straddling behavior are ``hotels and restaurants'' ($i=3$, straddles service-food border), ``manufacturing NEC, recycling'' ($i=7$, chemical-manufacturing), ``office, accounting, and computing machinery'' ($i=21$, manufacturing-service), ``aircraft and spacecraft'' ($i=22$, manufacturing-service), and ``research and development'' ($i=37$, manufacturing-service).

We also observe weak cogrouping at a larger scale, beyond that of single straddler industries. To study these grouping patterns, we use hierarchical clustering methods. We define the distance between industries to be
\begin{align}\label{distance_function}
d_{ij} = 1 - p_{ij}
\end{align}
where $p_{ij} \in [0,1]$ is the probability with which $i$ cogroups with $j$. To create a hierarchical tree, we use agglomerative clustering with the average linkage criterion. We find similar results using other distances and linkage criteria. We construct a tree by joining industries one-by-one, starting with the closest pair of industries and ending with the most separated. Distances between clusters of industries are defined as
\begin{align}
d_{AB}&= \frac{1}{|A| |B|}\sum_{i\in A, j\in B} d_{ij}\\
&= 1 - \frac{1}{|A| |B|}\sum_{i\in A, j\in B} p_{ij}\\
&= 1 - p_{AB},
\end{align}
where $p_{AB} \equiv \frac{1}{|A| |B|}\sum_{i\in A, j\in B} p_{ij}$ is the probability that a randomly picked pair from clusters $A$ and $B$ are cogrouped. This choice of cluster distance is known as the ``average linkage criterion'', and in the present context enables a simple interpretation of industry and cluster distances in terms of probabilities. In Appendix \ref{sec:overlap_distance} we discuss properties of the distance function Eq. \eqref{distance_function}.

The results of hierarchical clustering are shown in the dendrogram at the bottom of Fig. \ref{fig:coclassification_matrix}d. The dendrogram supports the five-way division into food, chemical, manufacturing, service, and energy industries. Further interpretation has to proceed cautiously, but we observe the following:
\begin{itemize}
\item The chemical and manufacturing industries appear to form a hierarchy in which the two communities are members of a larger ``chemo-manufacturing'' community.
\item Two large sub-communities appear within manufacturing. The industries in the upper left of the manufacturing block of Fig. \ref{fig:coclassification_matrix}d (7-10) are ``manufacturing NEC, recycling'', ``wood and products of wood and cork'', ``construction'', and ``other non-metallic mineral products'', and those in the bottom right (11-22) are various metal and machinery industries. The manufacturing group thus appears to divide into those industries that are structure-producing and those that are machinery-producing.
\item The machinery-producing industries further appear to contain two subsets. The first, industries 11-15, contains basic metal and machinery products. The second, industries 19-21, contains ``radio, television, and communication equipment'', ``medical, precision, and optical instruments'', and ``office, accounting, and computing machinery''. These industries appear to follow a ``precision equipment'' pattern. The four remaining machinery-producing industries that are not in either of these subsets ($i$=16-18,22) do not form their own cluster, but are all transportation equipment industries (ships and boats, motor vehicles, rail vehicles, aircraft and spacecraft).
\item The service community contains two well-connected subsets. One subset, ``health and social work'' and ``pharmaceuticals'' ($i$=35 \& 36), is health-oriented. The other subset is less clear cut; its seven members are ``finance, insurance'', ``post and telecommunications'', ``other business activities'', ``computer and related activities'', ``other community, social, and personal services'', ``education'', and ``pulp, paper, paper products, printing, and publishing''. Roughly, these sectors follow an ``information'' theme.
\end{itemize}

Although these groupings represent increased tendencies for intra-group transactions, the hierarchical structure given by the dendrogram in Fig. \ref{fig:coclassification_matrix} oversimplifies the community structure of the network somewhat. Hierarchical clustering forces hierarchical structure even where none exists \cite{Tibshirani2009}, and the actual clustering behavior may be more nuanced. The CCM displays substantial overlap between communities that is not apparent from the dendrogram in Fig. \ref{fig:coclassification_matrix}d. For example, the food and chemical industries show some tendency to cogroup; in certain countries (e.g. Australia) this cogrouping is strong. This behavior suggests an alternative hierarchy in which the two communities are members of a larger ``agrochemical'' community, or equivalently, overlap with the chemo-manufacturing community. As a second example, the service community as a whole shows overlap with part of the manufacturing community. The particular manufacturing industries overlapped tend to be ones further along the supply chain -- construction, radio, computer, medical, aircraft -- rather than basic materials industries -- metals, fabricated metal products, other non-metal materials. These particular manufacturing industries and the service industries may constitute some larger definition of the service community that includes its immediate suppliers.

It is also important to note that the communities at this level of aggregation are not mostly isolated clusters, but are more like perturbations on top of an otherwise strongly connected network. It is possible this behavior would change at lower levels of aggregation, with more narrow industry definitions serving to isolate industries from irrelevant parts of the economy.


\begin{figure*}[t!]
\includegraphics[width=\textwidth]{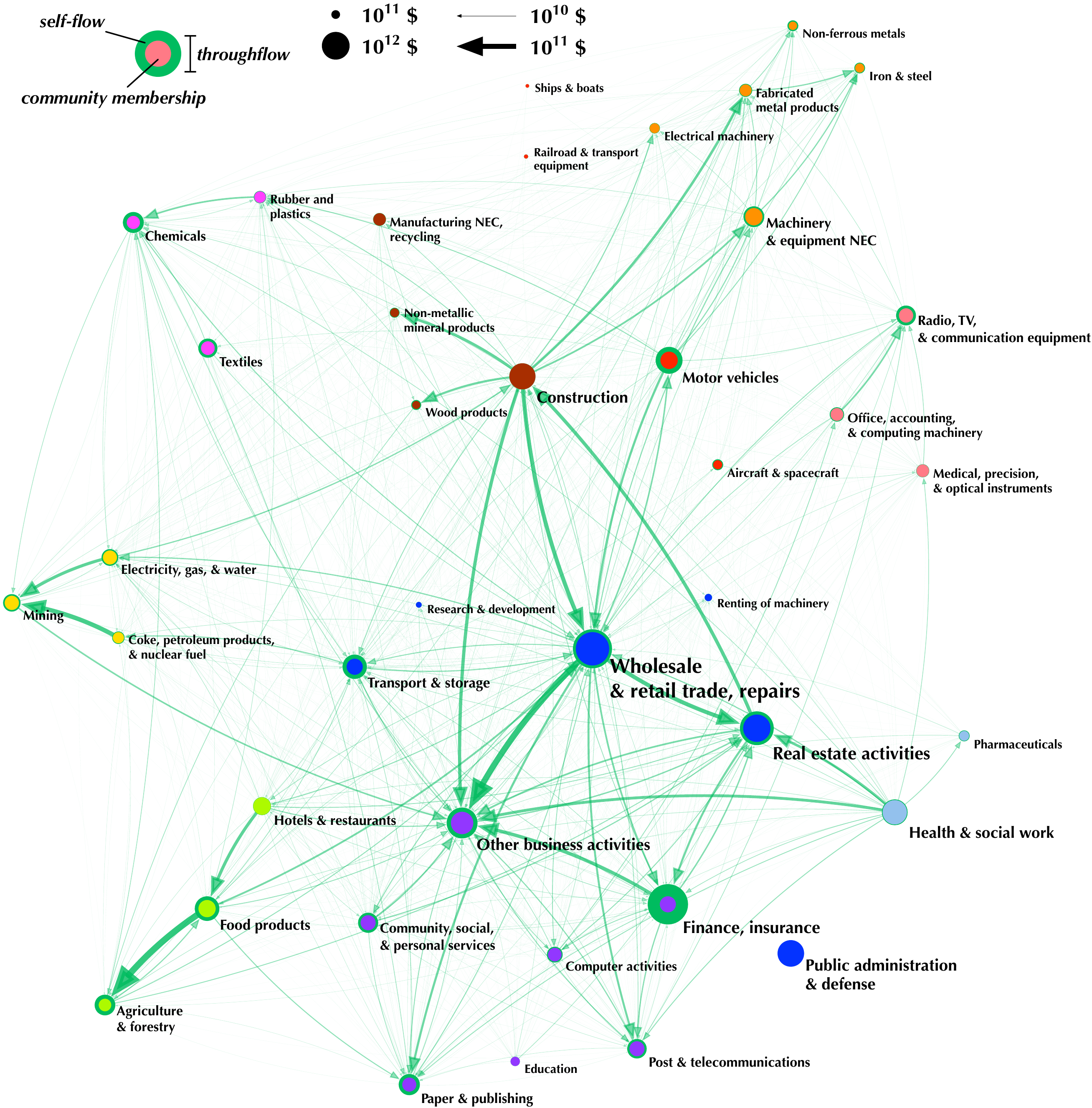}
\caption[Industry money flow network for the United States]{(Color online) The industry money flow network of the United States in 1997. Nodes are colored according to the communities identified in Fig. \ref{fig:coclassification_matrix}d. The size of a node corresponds to its throughflow (Eq. \eqref{throughflow}.) External flows $\vec{U}$ and $\vec{V}$ are omitted for clarity. To reduce picture file size, only flows larger than $\frac{1}{1000}$th of the largest flow are displayed. Remaining flows represent about 57\% of the $40^2=1600$ possible links. The true size of many of these flows can be best seen online by zooming in.  No intermediate consumption data was available for the ``Public administration and defense'' industry for the U.S, so it appears as an isolated node.}
\label{fig:US_IO_network}
\end{figure*}

\section{Discussion} \label{sec:discussion}
Comparisons of national economies typically focus on their differences; it is less often appreciated that economies may have substantial amounts of shared structure.  Chenery and Watanabe write, ``The structure of production, as defined by the input-output model, is the result of the interaction of a variety of forces, some leading to uniformity among countries and others to diversity. To the extent that production in various countries is intended to satisfy biologically determined human needs, is based on the same body of technological knowledge, and is constrained by the physical world, we should expect similarity in structure. To the extent that there are, among countries, variations in the relative scarcity of capital, labor and raw materials, differences in levels of income and composition of final demand, and variation in the scale of production, we may expect diversity.'' \cite{Chenery1958} While differences are apparent from statistics like GDP per capita or the export trade network, similarities are not yet well characterized. Such similarities can serve as constraints for theoretical and computational models of economies.
 
Both for the construction of such theories and further empirical work, the level of aggregation is important. Unlikely other networks where the meaning of a node is clear (as a person, city, router, web page, species, etc.), the meaning of nodes as industries is necessarily ambiguous and subject to arbitrary decisions on the part of the statistical agencies collecting economic data. These ambiguities are not drawbacks of the data per se, but rather reflect fundamental ambiguities in the distinctions between products, though they sometimes also reflect the limited resources of the statistical agencies. Because of this ambiguity, it is important for future theoretical and empirical work to account for the way results should change at different levels of aggregation.

A useful way to gauge the aggregation level of an industry network is to look at the amount of ``self-flow'' in the network. Self-flow represents transactions between firms that are classified within the same industry. Although these firms may produce different products, they are not different enough for them to have fallen into different industry bins. In this case, the industry partitioning scheme is too coarse-grained to differentiate them. The fraction of all intermediate flows that are self-flows, $\frac{\sum_i a_{ii}}{\sum_{jk} a_{jk}}$, can serve as a measure of the aggregation level of an industry network data set. For our data, this number varies between 0.15 to 0.30; that is, some 15 to 30\% of inter-industry money flows are really transactions of an industry with itself, reflecting the high level of aggregation of our data. Individual industries with large self-flows represent good candidates for subdivision in future I/O tables. (Table \ref{tab:industry_statistics}.)

\section{Conclusions}
Network methods are useful for studying the relationships between industries. Here, we have applied them to flows of money between industries. These networks are weighted, directed, dense, and contain self-links. We have characterized the flow weight and industry size distributions, identifying functional forms to serve as targets for theoretical models. We have examined the community structure of industries, finding groups corresponding to food, chemical, manufacturing, service, and energy industries, as well as nested sub-groups corresponding to finer categories of industries. Applying network methods to industrial money flows involves challenges not encountered in other network data sets, so to aid other researchers we have provided a brief introduction to the concepts and definitions of national accounting, as well as the measurement basis and interpretation of money flows.

\section{Acknowledgements}
JM gratefully acknowledge financial support from NSF Grant SBE0738187. We thank the International Institute for Applied Systems Analysis (IIASA) and the Young Scientist Summer Program (YSSP) where this research began, with financial support from The National Academies. We thank the Santa Fe Institute for support during the continuation of this research. We thank Aaron Clauset, Ben Good, and Doyne Farmer for several helpful conversations and suggestions.

\appendix

\section{Hierarchical clustering with the overlap distance}\label{sec:overlap_distance}
Define the overlap distance between nodes as 
\begin{align}
d_{ij} = 1 - p_{ij},
\end{align}
where $p_{ij}$ is the probability that nodes $i$ and $j$ are grouped in the same community. Since $p_{ij}$ is the probability that $i$ and $j$ are cogrouped, $d_{ij}$ is simply the probability that $i$ and $j$ are not cogrouped. To determine distances between clusters of nodes, let
\begin{align}
d_{AB}&= \frac{1}{|A| |B|}\sum_{i\in A, j\in B} d_{ij}\\
&= 1 - \frac{1}{|A| |B|}\sum_{i\in A, j\in B} p_{ij}\\
&= 1 - p_{AB},
\end{align}
where $p_{AB} \equiv \frac{1}{|A| |B|}\sum_{i\in A, j\in B} p_{ij}$ is the probability that a randomly picked pair from clusters $A$ and $B$ are cogrouped. This choice of $d_{AB}$ is known as the ``average linkage criterion''. In the present context it enables a simple interpretation of both node and cluster distances in terms of probabilities.

The elements $p_{ij}$ of the coclassification matrix cannot take on arbitrary values; the laws of probability impose interdependent constraints on matrix elements. Given the cogrouping probabilities $p_{ik}$ and $p_{jk}$ of $i$ and $j$ with some third node $k$, one can show that $p_{ij}$ is bound above and below as
\begin{align}
\max( 0, p_{ik} + p_{jk} - 1 ) \leq p_{ij} \leq 1 - \left| p_{ik} - p_{jk} \right|.
\end{align}
We can use these bounds to show two useful properties of the overlap distance. First, using the lower bound, one can show that
\begin{align}
d_{ij} \leq d_{ik} + d_{jk};
\end{align}
i.e. the overlap distance obeys triangle inequality.

Second, using the upper bound one may show that the overlap distance is equal to the Chebychev or $L^{\infty}$ distance applied to columns of the coclassification matrix:
\begin{align}
d_{ij} = \max_k \left| p_{ik} - p_{jk} \right|.
\end{align}
The $L^{\infty}$ distance is the largest absolute difference between elements of columns $i$ and $j$. Rewriting the upper bound as $\left|p_{ik} - p_{ij}\right| \leq 1 - p_{ij}$, we see that $\max_k \left| p_{ik} - p_{jk} \right|$ is at most $1 - p_{ij}$. To see that they are in fact equal, let $k=i$ and note that $\left| p_{ik} - p_{jk} \right| = \left| p_{ii} - p_{ji} \right| = 1- p_{ij}$. Since the argument of $\max_k$ achieves the largest possible value for at least one value of $k$, $d_{ij} = \max_k \left| p_{ik} - p_{jk} \right| = 1 - p_{ij}$.


\section{Industry flow statistics}
\begin{landscape}
\begin{table}[b!]
\caption[Industry flow statistics]{Industry statistics. All values are averages across countries. The thruflow column gives the average size of each industry $i$ as the average normalized throughflow $\bar{t_i} = \sum_c t_i^c/\sum_c 1$. (See Eq. \eqref{eq:normalized_thruflow}.) The sum over all industries sums to 100\%, within a small error due to rounding. The export, final, and intermediate revenues columns give the percentage of an industry's money in-flows received from each category. The three columns sum to 100\% for each industry. The imports, value added, and intermediate expenditure columns give the percentage of an industry's money out-flows going to each category. These three columns also sum to 100\% for each industry. The final column gives the self-flow $a_{ii}$ as a percentage of the total throughflow $t_i$ going through $i$.}
\rowcolors{2}{}{lightblue}
\small
\begin{tabular}{llc|ccc|ccc|c}
\hline\hline
\rowcolor{lightgray}
&&&\multicolumn{3}{c|}{\cellcolor{lightgray}\textbf{Money in-flows}}
&\multicolumn{3}{c|}{\cellcolor{lightgray}\textbf{Money out-flows}}&\\
\rowcolor{lightgray}
$\boldsymbol{i}$&
\textbf{Industry}&
\textbf{\% Thruflow}&
\textbf{\% Exports}&
\textbf{\% Final}&
\textbf{\% Int. rev.}&
\textbf{\% Imports}&
\textbf{\% Val. add.}&
\textbf{\% Int. exp.}&
\textbf{\% Self}
\rule[-2mm]{0pt}{4ex}\\
\hline
1	&Food products, beverages, and tobacco	&5.84	&16.8	&50.6	&32.6	&9.5	&25.0	&65.5	&14.0\\
2	&Agriculture, hunting, forestry, and fishing	&4.30	&10.9	&22.4	&66.7	&6.1	&50.6	&43.3	&13.1\\
3	&Hotels and restaurants	&2.88	&7.3	&69.9	&22.8	&4.7	&47.9	&47.4	&1.1\\
4	&Chemicals excluding pharmaceuticals	&2.35	&34.8	&12.5	&52.8	&20.0	&31.3	&48.7	&12.5\\
5	&Rubber and plastics products	&1.12	&25.3	&8.9	&65.8	&20.4	&35.0	&44.6	&6.9\\
6	&Textiles, textile products, leather, and footwear	&2.25	&32.7	&34.5	&32.8	&18.7	&33.4	&47.9	&18.6\\
7	&Manufacturing NEC, recycling	&1.07	&23.8	&48.3	&27.9	&12.8	&39.7	&47.5	&3.6\\
8	&Wood and products of wood and cork	&0.79	&18.4	&9.0	&72.6	&13.5	&33.7	&52.8	&14.6\\
9	&Construction	&7.68	&1.3	&76.5	&22.2	&7.5	&39.9	&52.6	&7.8\\
10	&Other non-metallic mineral products	&1.32	&16.5	&7.1	&76.4	&9.6	&39.3	&51.1	&9.7\\
11	&Fabricated metal products, except machin. and equip.	&1.87	&18.2	&15.0	&66.8	&12.6	&38.3	&49.2	&9.1\\
12	&Iron \& steel	&1.82	&33.0	&0.5	&66.5	&19.3	&25.8	&54.9	&19.2\\
13	&Machinery and equipment, NEC	&2.57	&33.3	&32.6	&34.2	&15.5	&37.8	&46.7	&7.9\\
14	&Electrical machinery and apparatus, NEC	&1.29	&36.8	&16.1	&47.1	&18.8	&35.1	&46.0	&7.5\\
15	&Non-ferrous metals	&0.48	&36.6	&2.3	&61.1	&24.7	&24.7	&50.6	&17.0\\
16	&Building and repairing of ships and boats	&0.42	&38.6	&34.0	&27.4	&16.0	&35.3	&48.7	&5.6\\
17	&Motor vehicles, trailers, and semi-trailers	&2.37	&38.1	&36.5	&25.4	&24.9	&25.1	&49.9	&14.1\\
18	&Railroad equipment and transport equipment, NEC	&0.25	&28.9	&38.5	&32.5	&19.7	&33.0	&47.4	&7.1\\
19	&Radio, television, and communication equipment	&0.98	&44.8	&29.8	&25.5	&25.7	&31.2	&43.1	&9.6\\
20	&Medical, precision, and optical instruments	&0.45	&36.9	&34.1	&29.0	&16.9	&40.3	&42.8	&4.2\\
21	&Office, accounting and computing machinery	&0.56	&54.3	&27.4	&18.3	&35.8	&26.1	&38.1	&3.8\\
22	&Aircraft and spacecraft	&0.23	&47.4	&16.9	&35.6	&27.7	&35.1	&37.2	&6.6\\
23	&Wholesale and retail trade, repairs 	&10.26	&7.7	&56.2	&36.1	&4.0	&57.0	&39.1	&4.0\\
24	&Finance, insurance 	&4.25	&4.8	&24.1	&71.1	&3.3	&60.1	&36.6	&12.9\\
25	&Post and telecommunications	&1.72	&5.1	&30.5	&64.5	&4.2	&66.7	&29.1	&4.6\\
26	&Other business activities	&4.34	&8.0	&14.4	&77.7	&4.3	&55.6	&40.1	&9.1\\
27	&Computer and related activities	&0.80	&6.1	&31.5	&62.4	&6.0	&54.4	&39.5	&5.4\\
28	&Other community, social, and personal services	&2.89	&3.8	&59.3	&36.9	&5.1	&53.0	&41.9	&7.7\\
29	&Education	&2.54	&0.6	&93.1	&6.2	&1.9	&75.7	&22.4	&0.8\\
30	&Pulp, paper, paper products, printing, and publishing	&2.54	&15.3	&16.7	&68.0	&13.5	&36.9	&49.6	&20.4\\
31	&Real estate activities	&6.03	&0.2	&73.8	&25.9	&1.1	&74.9	&24.0	&3.4\\
32	&Public admin. and defense; compulsory social sec.	&5.19	&0.5	&93.4	&6.0	&4.3	&64.2	&31.5	&1.3\\
33	&Renting of machinery and equipment	&0.44	&3.3	&19.8	&76.9	&3.0	&56.6	&40.4	&6.2\\
34	&Transport and storage	&4.84	&21.2	&26.9	&51.9	&9.3	&48.9	&41.8	&12.5\\
35	&Health and social work	&3.97	&0.4	&87.9	&11.6	&4.6	&62.6	&32.8	&4.0\\
36	&Pharmaceuticals 	&0.48	&25.4	&38.0	&36.6	&15.3	&37.8	&46.9	&6.6\\
37	&Research and development	&0.37	&8.0	&35.2	&56.8	&6.1	&56.6	&37.3	&3.7\\
38	&Mining and quarrying	&1.84	&21.4	&5.1	&73.5	&7.0	&55.4	&37.6	&4.7\\
39	&Coke, refined petroleum products, and nuclear fuel	&1.64	&20.9	&22.9	&56.1	&34.3	&20.0	&45.7	&6.0\\
40	&Electricity, gas, and water supply	&2.68	&1.6	&32.7	&65.7	&8.0	&50.4	&41.6	&12.2\\
\hline\hline
\end{tabular}
\label{tab:industry_statistics}
\end{table}
\end{landscape}


\begin{table*}
\center
\caption{Comparison of Weibull and lognormal fits to flow weight distribution.}
\rowcolors{2}{}{lightblue}
\begin{tabular}{l|cc|cc|cl}
\hline
\rowcolor{lightgray}
\textbf{Country}	&\multicolumn{2}{c|}{\cellcolor{lightgray}\textbf{Weibull}}		&\multicolumn{2}{c|}{\cellcolor{lightgray}\textbf{Lognormal}}		&\textbf{$\Delta $log-likelihood} &\textbf{Best fit}	\\
\rowcolor{lightgray}
	&$\lambda$	&$k$		&$m$		&$s$		&	&\\
\hline
Australia   &$8.81\times 10^{-5}$   &0.408   &-10.7   &2.96   &67.2   &Weibull\\
Brazil   &$1.88\times 10^{-4}$   &0.485   &-9.71   &2.44   &36.7   &Weibull\\
Canada   &$1.21\times 10^{-4}$   &0.483   &-10.1   &2.34   &2.99   &Weibull\\
China   &$1.58\times 10^{-4}$   &0.471   &-9.9   &2.44   &21.9   &Weibull\\
CzechRepublic   &$1.14\times 10^{-4}$   &0.471   &-10.2   &2.49   &48.4   &Weibull\\
Denmark   &$6.83\times 10^{-5}$   &0.433   &-10.8   &2.57   &-15.3   &lognormal\\
Finland   &$1.31\times 10^{-4}$   &0.489   &-9.99   &2.12   &-71.3   &lognormal\\
France   &$1.49\times 10^{-4}$   &0.518   &-9.86   &2.20   &12.3   &Weibull\\
Germany   &$1.44\times 10^{-4}$   &0.514   &-9.87   &2.12   &-30.1   &lognormal\\
Greece   &$4.14\times 10^{-5}$   &0.343   &-11.7   &3.45   &35   &Weibull\\
Hungary   &$1.34\times 10^{-4}$   &0.541   &-9.87   &1.94   &-63.4   &lognormal\\
Italy   &$1.29\times 10^{-4}$   &0.461   &-10.2   &2.56   &37   &Weibull\\
Japan   &$1.14\times 10^{-4}$   &0.423   &-10.5   &3.07   &124   &Weibull\\
Korea   &$1.07\times 10^{-4}$   &0.442   &-10.4   &2.58   &11.8   &Weibull\\
Netherlands   &$1.05\times 10^{-4}$   &0.494   &-10.2   &2.00   &-120   &lognormal\\
Norway   &$7.70\times 10^{-5}$   &0.463   &-10.6   &2.32   &-53.2   &lognormal\\
Poland   &$1.50\times 10^{-4}$   &0.490   &-9.88   &2.23   &-24.2   &lognormal\\
Spain   &$9.93\times 10^{-5}$   &0.469   &-10.3   &2.33   &-25.2   &lognormal\\
UnitedKingdom   &$8.31\times 10^{-5}$   &0.439   &-10.6   &2.63   &29.4   &Weibull\\
UnitedStates   &$9.23\times 10^{-5}$   &0.440   &-10.5   &2.52   &-26.2   &lognormal\\
pooled	&$1.08\times 10^{-4}$	&0.456	&-10.3	&2.54	&545.6	&Weibull\\
\hline
\end{tabular}
\label{tab:Weibull_v_lognormal}
\end{table*}

\bibliographystyle{elsarticle-num}
\bibliography{IOnet}
\end{document}